\documentstyle[prl,aps,twocolumn,floats,amssymb,epsfig]{revtex}
\def\bbC{{\mathbb C}}

\def\bbR{{\mathbb R}}

\def\su{{\rm{su}}}
\def\u{{\rm{u}}}
\def\SU{{\rm{SU}}}
\def\U{{\rm{U}}}

\def\SO{{\rm{SO}}}
\def\O{{\rm{O}}}

\newcommand{\mb}[1]{\ifmmode#1\else\mbox{$#1$}\fi}

\newcommand\la{\mb{\lambda}}

\newcommand\calG{\mb{{\cal G}}}
\newcommand\calH{\mb{{\cal H}}}

\newcommand\calL{\mb{{\cal L}}}
\newcommand\calM{\mb{{\cal M}}}
\newcommand\calN{\mb{{\cal N}}}

\newcommand\calV{\mb{{\cal V}}}

\newcommand{\beq}{\begin{equation}}
\newcommand{\eeq}{\end{equation}}
\newcommand{\nn}{\nonumber}
\newcommand{\bea}{\begin{eqnarray}}
\newcommand{\eea}{\end{eqnarray}}
\newcommand{\norm}[1]{\parallel \! {#1} \! \parallel}

\newcommand{\inprod}[2]{\langle {#1}, {#2} \rangle}

\newcommand{\inproda}[2]{\{ {#1}, {#2} \}}
\newcommand{\emb}{\mb{{\rm emb}}}
\newcommand{\asy}{\mb{{\rm asy}}}
\newcommand{\tr}{\mb{{\rm tr}\,}}

\newcommand{\pderiv}[2]{\frac{\partial {#1}}{\partial {#2}}}

\newcommand{\rhat}{\mb{\hat{\bm r}}}

\newcommand{\Ad}{{\rm Ad}}
\newcommand{\ad}{{\rm ad}}

\newcommand{\gsim}
{\raise.3ex\hbox{$\;>$\kern-.75em\lower1ex\hbox{$\sim$}$\:$}}
\newcommand{\lsim}
{\raise.3ex\hbox{$\;<$\kern-.75em\lower1ex\hbox{$\sim$}$\:$}}
\newcommand{\ts}{\textstyle}
\newcommand{\ds}{\displaystyle}
\newcommand{\half}{\mb{\ts \frac{1}{2}}}
\newcommand{\quarter}{\mb{\ts \frac{1}{4}}}

\newcommand{\bm}[1]{{\mbox{\boldmath $#1$}}}

\begin{document}
\renewcommand{\thesubsection}{\arabic{subsection}} % arabic subsection number
\draft

%%%%%%%%%%%%       abstract and title page    %%%%%%%%%%%%%%%%%%%%%%

%\preprint{}
\twocolumn[\hsize\textwidth\columnwidth\hsize\csname @twocolumnfalse\endcsname
\title{Asymptotically embedded defects}
\author{Nathan\  F.\  Lepora\footnotemark}
\date{Submitted 16 May, 2002; accepted 10 July 2002}
\maketitle

\begin{abstract}
For many cases, the conditions to fully embed a classical solution of one 
field theory within a larger theory cannot be met. Instead, we find it useful 
to embed only the solution's asymptotic fields as this relaxes the embedding 
constraints. Such asymptotically embedded defects have a simple
classification that can be used to construct classical solutions in general
field theories.
\end{abstract}
\pacs{Published in: Phys.\ Lett.\ B {\bf 541} (2002) 362.}
]

%%%%%%%%%%%%%         the text   %%%%%%%%%%%%%%%%%%%%%%%%%%%%%%%%%%%

\subsection{Introduction}

Embedded defects~\cite{vach94} are a useful way of
describing classical solutions to spontaneously broken Yang-Mills theories.
The idea is to take a defect in one symmetry breaking and then embed this
breaking in that of the full theory:
\bea
\label{1}
G\hspace{1em} &\rightarrow& \ H \nn\\
\cup\hspace{1em} && \hspace{.5em}\cup\\
G_\emb &\rightarrow& \ H_\emb.\nn
\eea
Then, if certain conditions are met, the embedded defect solves the 
full field equations.

A convenient feature of such embedded defects is that their construction
allows the degeneracy of the solutions to be easily found. Then the
embedded classification is complementary to the topological 
classification~\cite{tom}, where only the existence of the classical 
solutions is usually clear. Degeneracies of classical solutions are
relevant to non-Abelian duality~\cite{gno}, the collective coordinate
quantization~\cite{coll}, zero modes and the global properties of
gauge transformations~\cite{globcol}.

The construction of embedded defects
has been closely examined for vortices~\cite{mevor}
and monopoles~\cite{memon}. 
For both cases, one of the embedding conditions 
gives a geometric constraint on (\ref{1}). This constraint is simply stated 
in terms of the Lie algebras of the two theories.

In addition to this geometric constraint, there is another condition on the
fields of the embedding. This condition
was shown, in reference~\cite{vach94}, to hold when
\beq
\label{2}
D(G_\emb)\calV_\emb=\calV_\emb
\eeq
with $D$ the representation and $\calV_\emb$ the embedded vector space of
scalar field values. Essentially, (\ref{2})
constrains the pairs $(\calV_\emb,G_\emb)$ that define the embedded theory.
However, when examining embedded monopoles~\cite{memon}, we found
condition (\ref{2}) to be so constraining that it only holds in the
simplest cases. Because of this, we instead considered
asymptotically embedded monopoles, which are only embedded at
infinity and can behave differently in the core.
Such asymptotically embedded monopoles are not constrained by (\ref{2}).

The aim of this paper is to generalize this idea of asymptotic embedding 
to other defects and then study how these defects are constrained. 
Just like embedded monopoles, we find that many defects can only be
asymptotically embedded and, then, the problematic condition 
(\ref{2}) does not apply.

The plan of this paper is as follows. In section~\ref{sec2}, 
we review the theory
of embedded defects, then in section~\ref{sec3} we discuss how the embedding 
(\ref{1}) is constrained --- generalizing the discussion from
references~\cite{mevor,memon}. After giving an example of a non-embedded vortex
in section~\ref{sec4}, we consider asymptotically embedded defects and their
properties in section~\ref{sec5}. 
To conclude, we summarize our main results in section~\ref{sec6}.

\subsection{Embedded defects}
\label{sec2}

Consider a spontaneously broken Yang-Mills theory 
with compact gauge group $G$ and a scalar field $\Phi\in \calV$
in the $D$ representation of $G$
\bea  
\label{lag}
\calL[\Phi,A^\mu] &=& -\quarter \inprod{F_{\mu\nu}}{F^{\mu\nu}} + 
\half \inprod{D_\mu\Phi}{D^\mu\Phi} - V[\Phi],\\
&&F^{\mu\nu} = \partial^\mu A^\nu - \partial^\nu A^\mu + [A^\mu,A^\nu],\\
&&D^\mu\Phi = \partial^\mu\Phi + d(A^\mu)\Phi.
\eea
In this Lagrangian density, the derived representation $d(X)$
is defined by $e^{d(X)}=D(e^X)$ with $X$ in the Lie algebra $\calG$.
Then $\calL[\Phi,A^\mu]$ is invariant under the gauge transformation
\beq
\label{gt}
\Phi\mapsto D(g)\Phi,\hspace{1em}
A^\mu\mapsto\Ad(g)A^\mu - (\partial^\mu g)g^{-1}
\eeq
with $g(x)\in G$ and $x=(\bm r,t)$.

If the potential $V[\Phi]$ is minimized at some value $\Phi_0$,
the residual gauge symmetry is
\beq
H = \left\{h\in G:D(h)\Phi_0=\Phi_0\right\},
\eeq
where $\calH$ is the Lie algebra of $H$.
This defines an $\Ad(H)$-invariant decomposition of $\calG$
into massless and massive gauge boson generators
\beq
\label{reddec}
\calG = \calH \oplus \calM,\hspace{2em}
\Ad(H)\calH \subseteq \calH,\hspace{.5em}
\Ad(H)\calM \subseteq \calM.
\eeq
Here $\Ad$ refers to the adjoint representation of $G$ on $\calG$,
where $\Ad(g)X=gXg^{-1}$ and the derived representation is
$\ad(X)Y=[X,Y]$ for $X,Y\in\calG$.

In this paper, we use a coordinate independent notation, which
we believe better reflects the geometry behind most of our
results~\cite{mevor,megeom}. 
Then the gauge kinetic term in (\ref{lag}) is
defined by an inner-product on $\calG$
\beq
\label{XY}
\inprod{X}{Y} = \frac{1}{f_a^2}\inproda{X}{Y}_a,\hspace{2em}
X,Y\in\calG_a
\eeq
with $f_a$ a coupling constant and $\inproda{X}{Y}_a$ a real, 
$\Ad(G_a)$-invariant inner-product on each simple or $\u(1)$ subgroup 
$\calG_a\subseteq\calG$~\cite{com2}.
Likewise, the scalar kinetic term is defined by a Euclidean inner-product 
$\inprod{\Phi_1}{\Phi_2}$ on $\calV$~\cite{com}.

Then, in this notation, the field equations are
\bea
\label{fe}
D_\mu D^\mu\Phi=-\pderiv{V}{\Phi},\hspace{1.5em}
D_\mu F^{\mu\nu}=J^\nu,
\eea
where the current and covariant derivative are
\bea
\label{Jnu}
\inprod{J^\nu}{Y}&=&\inprod{d(Y)\Phi}{D^\nu\Phi}-\inprod{D^\nu\Phi}{d(Y)\Phi},
\\
D_\mu F^{\mu\nu}&=&\partial_\mu F^{\mu\nu}+[A_\mu, F^{\mu\nu}].
\eea

Embedded defects~\cite{vach94}
are $(2-k)$-dimensional defects that remain solutions
when embedded into a larger gauge theory
(so $k=0,1,2$ for domain walls, vortices and monopoles). 
They are defined by an embedding of their gauge symmetry breaking in
that of the full theory
\bea
\label{emb}
G\hspace{1em} &\rightarrow& \ H \nn\\
\cup\hspace{1em} && \hspace{.5em}\cup\\
G_\emb &\rightarrow& \ H_\emb. \nn
\eea
Such embedded defects have fields fully embedded over their
space-time domain
\bea
\label{embform}
\Phi_\emb(\bm r,t)\in\calV_\emb,\hspace{.5em}
A^\mu_\emb(\bm r,t)\in\calG_\emb,\hspace{.8em}\bm r\in\bbR^{1+k},
\eea
where $\calV_\emb\subseteq\calV$ is a vector subspace. Note that
it will often be useful to embed topological defects, which have
$\pi_k (G_\emb / H_\emb)\neq 0$.

In reference~\cite{vach94} another condition was given for a defect to 
be embedded:
\beq
\label{DG}
D(G_\emb)\calV_\emb=\calV_\emb,
\eeq
because the constraint in (b), below, is then satisfied. Later in this
paper we find this condition to be so constraining that it only
holds for the simplest cases. This motivates our discussion of
asymptotically embedded defects, for which condition (\ref{DG}) is relaxed.

An embedded defect is a solution of the full theory if the field
equations reduce to consistent field equations 
on its embedded theory~\cite{vach94}. This gives four constraints 
from the two field equations in (\ref{fe}):\\
\noindent (a) The current, found from $\Phi_\emb$ and $A^\mu_\emb$,
satisfies
\beq
\label{c1}
\inprod{J^\nu\!}{\,\calG_\emb^\perp}=0,
\eeq
where $\calG=\calG_\emb\oplus\calG_\emb^\perp$.
This constrains how $\calG_\emb\subseteq\calG$.\\
\noindent (b) The kinetic scalar term satisfies
\beq
\label{c2}
\inprod{D_\mu D^\mu \Phi_\emb}{\calV_\emb^\perp}=0,
\eeq
where $\calV=\calV_\emb\oplus\calV_\emb^\perp$. As mentioned above, this 
always holds when (\ref{DG}) is satisfied~\cite{vach94}.\\
\noindent (c) The scalar potential, found from $\Phi_\emb$, satisfies
\bea
\label{scderiv}
\inprod{\pderiv{V}{\Phi}}{\calV_\emb^\perp}=0.
\label{eq-2}
\eea
This constrains the potential --- for example, it holds in the 
BPS limit.\\
\noindent (d) The gauge kinetic terms satisfies
\beq
\label{DF}
\inprod{D_\mu F^{\mu\nu}}{\calG_\emb^\perp}=0.
\eeq
This always holds by algebraic closure of $\calG_\emb$~\cite{vach94}.

\subsection{Constraining the spectrum of embedded defects}
\label{sec3}

The condition $\inprod{J^\nu\!}{\,\calG_\emb^\perp}=0$ in (a) above has
been shown to constrain the spectrum of embedded vortices~\cite{mevor} 
and embedded monopoles~\cite{memon}. Here we give a general proof of this
constraint for embedded defects.

The following proof relies on
reducing $\calM$ into irreducible subspaces under the adjoint action 
of $H$. These correspond to irreducible representations of $H$ 
on $\calM$:
\beq
\label{Mdecom}
\calM= \calM_1 \oplus \cdots \oplus \calM_n,\hspace{1.5em}
\Ad(H)\calM_a\subseteq\calM_a.
\eeq
Physically, each $\calM_a$ defines a gauge multiplet of 
massive gauge bosons (for example, the W- and Z-bosons in electroweak theory).

To examine condition (\ref{c1}), we firstly rewrite (\ref{emb}) as
\bea
\label{decomps}
\begin{array}{ccccc}
\calG & = & \calH & \oplus\ & \calM \\
\cup && \cup && \cup \\
\calG_\emb & = & \calH_\emb & \oplus & \calN
\end{array}
\eea
and split $\calN$ into its $\Ad(H_\emb)$-irreducible subspaces
\beq
\label{Ndecom}
\calN=\calN_1 \oplus \cdots \oplus \calN_m,\hspace{1.5em}
\Ad(H_\emb)\calN_a\subseteq\calN_a
\eeq
in a similar way to $\calM$ in (\ref{Mdecom}).
It will also be useful to consider an embedded defect with scalar field 
generated by the action of $G$ upon $\Phi_0$~\cite{noteans}
\bea
\label{23}
\Phi_\emb(x) = h(x)D[g(x)]\Phi_0,\hspace{1em}g(x) \in G
\eea
with $h(x)$ a function of $x=(\bm r,t)$. Consequently,
a gauge transformation (\ref{gt}) with group element $g^{-1}(x)$
takes the embedded defect to a unitary gauge~\cite{uncom}
\bea
\label{unans}
\Phi_\emb = h(x)\Phi_0,\hspace{1em}
A^\mu_\emb =C_\emb^\mu(x) + W_\emb^\mu(x),
\eea
where the gauge field naturally separates into massless
$C^\mu_\emb\in\calH_\emb$ and massive $W^\mu_\emb\in\calN$ parts. 

Then a substitution of $J^\nu$ from (\ref{Jnu})
into (\ref{c1}) constrains the ansatz (\ref{unans}) by
\beq
\label{25}
\inprod{d(\calG_\emb^\perp)\Phi_0}{D^\nu\Phi}
=\inprod{D^\nu\Phi}{d(\calG_\emb^\perp)\Phi}=0.
\eeq
Now, in a unitary gauge
\beq
D^\nu\Phi=(\partial^\nu h)\Phi_0 + h\,d(W_\emb^\nu)\Phi_0.
\eeq
Using this and noting $\inprod{d(\calG)\Phi_0}{\Phi_0}=0$, we see that
(\ref{25}) holds when
\beq
\label{27}
\inprod{d(\calG_\emb^\perp)\Phi_0}{d(\calN)\Phi_0}
=\inprod{d(\calN)\Phi_0}{d(\calG_\emb^\perp)\Phi_0}=0,
\eeq
which algebraically constrains the embedding (\ref{decomps}).

This algebraic constraint (\ref{27}) is expressed more simply
by applying the following result~\cite{mevor}:
\bea
\inprod{d(X_a)\Phi_0}{d(Y_b)\Phi_0} =
\la_a\la_b\inprod{X_a}{Y_b},\hspace{2em}\nn\\
\la_a=\frac{\norm{d(X_a)\Phi_0}}{\norm{X_a}},\hspace{.5em}
X_a\in \calM_a,Y_b\in\calM_b,
\eea
which says the representation respects orthogonality between the 
irreducible subspaces $\calM_a$ in (\ref{Mdecom}). 
Therefore when $\la_a\neq\la_b$, we have
\beq
\label{Nsub}
\calN_c \subseteq \calM_a
\eeq
with $\calN_c$ a subspace from (\ref{Ndecom}).
If $\la_a=\la_b$, the embedding can also be between gauge families 
(see reference~\cite{mevor}).

When $\calN$ is irreducible under $\Ad(H_\emb)$, so (\ref{Ndecom}) is trivial,
the constraint (\ref{Nsub}) becomes more simply $\calN\subseteq\calM_a$.
This is the case with embedded vortices and embedded monopoles, as we now
discuss.

For embedded vortices, which are Nielsen-Olesen vortices~\cite{nol} embedded
like (\ref{emb}), the embedding is~\cite{noteans}
\bea
\label{voremb}
\begin{array}{ccc}
\calG &\rightarrow& \calH \\
\cup && \cup \\
\u(1) &\rightarrow& 0
\end{array}
\eea
with $\U(1)=\exp(X\vartheta)$ and $X\in\calM$. In the radial
gauge, the embedded ansatz (\ref{embform}) is
\bea
\Phi_\emb(r,\vartheta) =\! f(r)D(e^{X\vartheta})\Phi_0,\hspace{.3em}
\bm A_\emb(r,\vartheta) =\! \ds\frac{g(r)}{r}X\hat\bm\vartheta.\!
\eea
In the unitary gauge, this ansatz becomes~\cite{uncom}
\bea
\Phi_\emb(r,\vartheta) = f(r)\Phi_0,\hspace{1em}
\bm A_\emb(r,\vartheta) = \ds\frac{g(r)+1}{r}X\hat\bm\vartheta.
\eea
Then (\ref{Nsub}) constrains the vortex's embedding by~\cite{mevor}
\beq
\calN \subseteq \calM_a,\hspace{1em}\calN=\bbR X
\eeq
with $\calM_a$ an irreducible subspace from (\ref{Mdecom}).

For embedded monopoles, which are 't~Hooft-Polyakov monopoles~\cite{hooft}
embedded like (\ref{emb}), the embedding is
\bea
\begin{array}{ccc}
\calG &\rightarrow& \calH \\
\cup && \cup \\
\su(2) &\rightarrow& \u(1)
\end{array}
\eea
with $\su(2)$ basis $[t_a,t_b]=\epsilon_{abc}t_c$ and
$\U(1)=\exp(t_3\vartheta)$. In the radial gauge, the embedded ansatz 
(\ref{embform}) is
\bea
\Phi(\bm r) = \frac{H}{r}\,D[g(\rhat)]\Phi_0,\hspace{1.2em}
A^i(\bm r) = \ds\frac{K-1}{r}\,\epsilon_{iab}\hat x_bt_a,
\eea
where $g(\rhat)=e^{\varphi t_3}e^{\vartheta t_2}e^{-\varphi t_3}$ in
spherical polars. In the unitary gauge, this ansatz becomes~\cite{uncom,jr}
\bea
\label{ungauge}
\Phi(\bm r)=\frac{H}{r}\,\Phi_0,\hspace{1.5em}
\bm A(\bm r)=-\bm A_D\,t_3 -\frac{K}{r}\,\hat\bm\eta_st_s
\eea
with $\bm A_D=\hat\bm\varphi(1-\cos\vartheta)/r\sin\vartheta$ the 
Dirac gauge potential ($\bm\nabla\cdot\bm A_D=0$,
$\bm\nabla\wedge\bm A_D=\rhat/r^2$) and
\beq
\hat\bm\eta_1=\sin\varphi\,\hat\bm\vartheta
+\cos\varphi\,\hat\bm\varphi,\hspace{1em}
\hat\bm\eta_2=-\cos\varphi\,\hat\bm\vartheta
+\sin\varphi\,\hat\bm\varphi
\eeq
are two orthonormal unit-vectors orthogonal to $\rhat$. 
Then (\ref{Nsub}) constrains the monopole's embedding by~\cite{memon}
\beq
\calN \subseteq \calM_a,\hspace{1em}\calN=\bbR t_1 + \bbR t_2,
\eeq
where $\calM_a$ is an irreducible subspace from (\ref{Mdecom}).

\subsection{An illustrative example: vortices in $\bm{\SU(3)\rightarrow\U(2)}$}
\label{sec4}

Let us consider a spontaneously broken $\SU(3)$ Yang-Mills theory with
scalar field in the adjoint representation. This theory should give a typical 
example of a non-Abelian gauge theory in which to construct embedded defects.

It will be convenient to define the following diagonal generators
\beq
T_1=i\,{\rm diag}(1,1,-2),\hspace{1em}
T_2=i\,{\rm diag}(1,-1,0).
\eeq
Then a vacuum $\Phi_0=v\, T_1$ gives a residual gauge symmetry $H=\U(2)$
that satisfies $\Ad(H)\Phi_0=\Phi_0$ with Lie algebra
\bea
\calH=\left(\begin{array}{ccc} \su(2)&\vdots&\bm 0 \\ 
\cdots&\cdots&\cdots\\ \bm 0^\dagger&\vdots&0\end{array}\right)
\oplus \u(1)_{T_1},\hspace{1em}
u(1)_{T_1}=\bbR\,T_1.
\eea
This algebra defines a split $\su(3)=\calH\oplus\calM$, where
\bea
\calM=\{X(\bm v):\bm v \in\bbC^2\},\hspace{.6em} X(\bm v)=
\left(\begin{array}{ccc} \bm 0_2&\vdots&\bm v \\ 
\cdots&\cdots&\cdots\\ \!-\bm v^\dagger\!\!\!&\vdots&0\end{array}\right)\!\!
\eea
with $\calM$ irreducible under $\Ad(H)$.

Using the methods in section~\ref{sec2}, we consider an embedded vortex ansatz
in a radial gauge
\bea
\label{vorans}
\Phi_\emb=f(r)\Ad(e^{\vartheta X_1})\Phi_0,\hspace{1em}
\bm A_\emb=\frac{g(r)}{r}X_1\hat\bm\vartheta
\eea
with generator $X_1=X(\half\bm e_1)$, where $\bm e_1^\dagger=(1,0)$. 
Then $\calV_\emb$, the vector space of $\Phi_\emb$ values, is
constructed by expanding
\bea
\Ad(e^{X_1\vartheta})\Phi_0&=&\Phi_0+\vartheta\,\ad(X_1)\Phi_0 \nn\\
&&\hspace{2em}+ \half\vartheta^2\,\ad(X_1)\ad(X_1)\Phi_0 + \cdots
\eea
In this expansion
\beq
\label{exam}
\ad(X_1)\Phi_0=-3vY_1,\hspace{1em}
\ad(X_1)\ad(X_1)\Phi_0=-9vX_1,
\eeq
where $Y_1=X(\half i\bm e_1)$. Therefore $\calV_\emb$ is spanned by at
least three independent generators $T_1, X_1, Y_1$.

However, (\ref{vorans}) is supposed to be an embedded vortex, for
which $\calV_\emb$ should be isomorphic to either $\bbC$ 
or $\bbR^2$ in a radial gauge. 
Therefore we conclude that (\ref{vorans}) is not an embedded vortex. 

Another way to see that (\ref{vorans}) is not an embedded vortex is by
calculating
\bea
\Ad(e^{\vartheta X_1})T_1&=&
\left(\ts\frac{3}{4}\cos\vartheta+\quarter\right)T_1\nn\\&&+\,
\left(\ts\frac{3}{4}\cos\vartheta-\ts\frac{3}{4}\right)T_2-
\ts\frac{3}{2}\sin\vartheta\,Y_1.
\eea
Evidently, this forms a complicated curve that is not contained within
a two-dimensional vector space.

\subsection{Asymptotically embedded defects}
\label{sec5}

Essentially, the vortex in section~\ref{sec4} is not embedded because the 
condition $D(G_\emb)\calV_\emb=\calV_\emb$ does not hold for any 
$\calV_\emb\cong \bbR^2$. However, the simple classification in 
section~\ref{sec3}
still applies, so let us see if we can generalize the
definition of an embedded defect in a useful way.

For this definition, we consider an {\em asymptotically embedded defect} to have
asymptotic fields
\beq
\Phi_\asy\sim\Phi_\emb(x)\in\calV_\emb,\hspace{1em}
A^\mu_\asy\sim A^\mu_\emb(x)\in\calG_\emb.
\eeq
These are asymptotically similar to an embedded defect, but can differ
from that form elsewhere. We also suppose that the scalar field is
like (\ref{23}), so in a unitary gauge
\bea
\label{ga}
\Phi_\asy \sim h(x)\Phi_0,\hspace{1em}
\bm A_\asy^\mu \sim C_\emb^\mu(x) + W^\mu_\emb(x)
\eea
with $h(x)\rightarrow 1$ as $r\rightarrow\infty$.

To simplify matters, we take $h=h(r)$ to be a radial function
and suppose the gauge field satisfies
\beq
\partial_\mu W^\mu_\emb=0,\hspace{1em}\hat x^iW^i_\emb=0.
\eeq
All of the above conditions are motivated by the properties of embedded 
vortices and embedded monopoles.

Now, we examine condition (b) in section~\ref{sec2}, which requires
$\inprod{D_\mu D^\mu \Phi_\emb}{\calV_\emb^\perp}=0$. In this condition,
the second-order covariant derivative is
\bea
\label{eDD}
D_\mu D^\mu\Phi_\emb&=& 
(\nabla^2h)\,\Phi_0 +h\,g_{\mu\nu}d(W^\mu_\emb)d(W^\nu_\emb)\Phi_0\nn\\
&&\hspace{.2em}+\,h\,d(\partial_\mu W^\mu_\emb)\Phi_0
- \partial^i h\,d(W^i_\emb)\Phi_0.
\eea
Since $h=h(r)$ is radial, then $\partial^ih=h'\hat r^i$ and so by (\ref{ga})
\bea
\label{DD}
D_\mu D^\mu\Phi_\emb= (\nabla^2h)\,\Phi_0 +hg_{\mu\nu}
d(W^\mu_\emb)d(W^\nu_\emb)\Phi_0.\!
\eea
Then condition (b) in section~\ref{sec2} holds when
\beq
\label{dd}
g_{\mu\nu}d(W^\mu_\emb)d(W^\nu_\emb)\Phi_0 \propto \Phi_0.
\eeq
This is very constraining on both the defect's fields
and embedding $\calN\subset\calM$.

Therefore when both conditions (\ref{Nsub}) and (\ref{dd}) hold, there are
embedded defect solutions like (\ref{embform}). These solve the full field
equations.

Alternatively, if (\ref{dd}) does not hold,
the defect may then be asymptotically embedded. This situation 
happens when the second 
term of (\ref{DD}) is negligible compared to the first; namely, when 
asymptotically
\beq
\label{asycond}
\O(\nabla^2h_{\rm asy}) > \O(W^2_{\rm asy}).
\eeq
Whether this is the case depends on the nature of the defect and 
the parameters of the theory. Such asymptotically embedded defects
are constrained only by the analysis in section~\ref{sec3}, which requires
$\calN_a\subseteq\calM_b$.

To illustrate these results, we now consider
embedded vortices and embedded monopoles.

For embedded vortices, condition (\ref{dd}) becomes
\beq
\label{vordd}
d(X)d(X)\Phi_0\propto\Phi_0,
\eeq
where $X$ is the vortex generator in (\ref{vorans}). This equation is very 
constraining on
$X$ and does not always hold [for example, see equation (\ref{exam})
in section~\ref{sec4}]. When (\ref{vordd}) does not hold,
the vortex could instead be asymptotically embedded; this
depends on the scalar and gauge masses in the embedded theory, 
$m_S$ and $m_V$, that define the asymptotic Nielsen-Olesen 
profiles~\cite{nol,tommark}
\bea
\label{vorprof}
f(r)-v&=&
\left\{ 
\begin{array}{ll}
\O[r^{-1/2}\exp(-m_Sr)], & m_S\leq 2m_V, \\
\O[r^{-1}\exp(-2m_Vr)], & m_S\geq 2m_V, 
\end{array}\right. \\
g(r)+1&=&\O[r^{1/2}\exp(-m_Vr)].
\eea
Therefore the vortex can be asymptotically embedded when $m_S<2m_V$.

For embedded monopoles, condition (\ref{dd}) becomes
\beq
\label{mondd}
d(t_1)d(t_1)\Phi_0+d(t_2)d(t_2)\Phi_0\propto\Phi_0.
\eeq
Again this is very constraining and does not generally hold. When
(\ref{mondd}) is not the case, the monopole may instead be asymptotically 
embedded; this depends upon the asymptotic 't~Hooft-Polyakov profiles
\bea
H(r)-v&=&\O[\exp(-m_S r)],\\
K(r) &=&\O[\exp(-m_Vr)].
\eea
Therefore the monopole can be asymptotically embedded when $m_S<2m_V$, as 
found in reference~\cite{memon}.

To finish this section, we discuss a couple of points about asymptotically
embedded vortices and monopoles.

The first point is that condition (\ref{mondd}) essentially
generalizes (\ref{vordd}) to two generators. This generalization
relates to the property that any monopole embedding 
contains vortex embeddings:
\beq
\label{monvor}
\begin{array}{ccccc}
\u(1) & \subseteq & \su(2)& \subseteq & \calG \\
\downarrow && \downarrow && \downarrow \\
0 & \subseteq & \u(1) & \subseteq & \calH
\end{array}
\eeq
For further discussion, we refer to reference~\cite{memon}.

The other point is that conditions (\ref{vordd}) and (\ref{DG}) are the same
in many situations. For example, the relation $d(X)d(X)\Phi_0=c^2\Phi_0$
is the same as
\beq
\label{e1}
D(e^{X\theta})\Phi_0=\cos\theta\,\Phi_0+\sin\theta\,d(X/c)\Phi_0.
\eeq
Then (\ref{e1}) means that
$\calV_\emb=\bbR\,D(e^{X\theta})\Phi_0$ is a two-dimensional vector
space with $D(\U(1)_\emb)\calV_\emb=\calV_\emb$. Thus
(\ref{vordd}) and (\ref{DG}) are equivalent for embedded vortices.

In this paper, however, we find the constraint (\ref{vordd}) more 
convenient because it more clearly relates to the definition
of asymptotically embedded defects.

\subsection{Conclusions}
\label{sec6}

To conclude, we summarize our results and make a comment.

\noindent (a) Our arguments relate to the following decomposition
%decomposition of a symmetry breaking $G\rightarrow H$
\[
\calG=\calH\oplus\calM,\hspace{2em}
\calM=\calM_1\oplus\cdots\oplus\calM_n
\]
with each $\calM_a$ irreducible under $\Ad(H)$.

\noindent (b) Then defect embeddings are constructed by
\[
\begin{array}{ccccc}
\calG & = & \calH & \oplus\ & \calM \\
\cup && \cup && \cup \\
\calG_\emb & = & \calH_\emb & \oplus & \calN
\end{array}
\]
and satisfy a constraint
\[
\calN_a \subset \calM_b,
\]
where $\calN=\calN_1\oplus\cdots\oplus\calN_m$ under $\Ad(H_\emb)$.

\noindent(c)
Whether defects are fully embedded or asymptotically
embedded depends on whether the embedded ansatz
\[
\Phi=h(r)\Phi_0,\hspace{1em}
A^\mu_\emb=C^\mu_\emb + W^\mu_\emb
\]
satisfies the field equations everywhere or only asymptotically.
The defect is fully embedded when 
\[
g_{\mu\nu}d(W_\emb^\mu)d(W_\emb^\nu)\Phi_0\propto\Phi_0
\]
(with certain conditions on $W^\mu_\emb$); otherwise, the defect is 
asymptotically embedded when
\[
\O(\nabla^2h_{\rm asy})<O(W^2_{\rm asy}),
\]
which depends on the scalar and gauge masses within the embedded theory, 
$m_S$ and $m_V$. 
Such asymptotically embedded defects are constrained only by (b).

\noindent (d)
For embedded Nielsen-Olesen vortices
\[
\Phi_\emb=f(r)\Ad(e^{\vartheta X_1})\Phi_0,\hspace{1em}
\bm A_\emb=\frac{g(r)}{r}X_1\hat\bm\vartheta,
\]
the vortex embedding is therefore constrained by
\[
\calN\in\calM_a,\hspace{1em}\calN=\bbR\, X.
\]
Furthermore, the vortex is fully embedded when
\[
d(X)d(X)\Phi_0\propto\Phi_0
\]
and, otherwise, asymptotically embedded if $m_S<2m_V$.

\noindent (e)
For embedded 't~Hooft-Polyakov monopoles
\[
\Phi(\bm r) = \frac{H}{r}\,D(g(\rhat))\Phi_0,\hspace{1.2em}
A^i(\bm r) = \ds\frac{K-1}{r}\,\epsilon_{iab}\hat x_bt_a,
\]
the monopole embedding is therefore constrained by
\[
\calN\in\calM_a,\hspace{1em}\calN=\bbR\,t_1+\bbR\,t_2.
\]
Furthermore, the monopole is fully embedded when
\[
d(t_1)d(t_1)\Phi_0+d(t_2)d(t_2)\Phi_0\propto\Phi_0
\]
and, otherwise, asymptotically embedded if $m_S<2m_V$.

\noindent (f) 
For a final comment, we note that the formalism in (a) to (c) is
general. Therefore more complicated embeddings can be constructed
--- for example, electroweak theory in flipped-$\SU(5)$
\bea
\begin{array}{ccc}
\su(5)\oplus\u(1)&\rightarrow&\su(3)\oplus\su(2)\oplus\u(1)\nn\\
\cup && \cup \nn\\
\su(2)\oplus\u(1)&\rightarrow&\u(1).\nn
\end{array}
\eea
Such an embedding could be useful for considering the counterparts of 
Z-strings, W-strings, sphalerons and Nambu 
monopoles in a grand unified theory.

%%%%%%%%%%%%%%%%%%%%%%%%%%%  bibliography   %%%%%%%%%%%%%%%%%%%%%%%

\footnotetext[1]{\ email address: n$\_$lepora@hotmail.com}


\begin{thebibliography}{99}
\bibitem{com} We have used same notation for both inner-products because the
one referred to should be clear from the context.
\bibitem{noteans}
Note there are more general ansatz\"e with solutions not generated
from a single value $\Phi_0$ --- for example, see 
reference~\cite{tommark} for a vortex with such an ansatz in 
$\SO(3)\rightarrow\bm 1$.
\bibitem{uncom}
While the field equations and embedding conditions are equivalent in the
radial and unitary gauges, care should be taken with the 
stability analysis. This is because the physical fluctuations are 
non-singular in the radial gauge, but can be gauge transformed to singular 
fluctuations in the unitary gauge
\end{thebibliography}
\end{document}